\documentclass[fleqn,twoside]{article}
\usepackage{espcrc2}
\usepackage{graphicx}

\begin{document}

\title{Improved Measure of Local Chirality%
\thanks{Presented by T.\ Draper at Lattice 2004.}%
\thanks{This work is supported in part by the U.S. Department of Energy
        under grant numbers DE-FG05-84ER40154 and DE-FG02-02ER45967.}
}
\author{%
Terrence Draper
  \address[UK]{Department of Physics and Astronomy, 
               University of Kentucky, 
               Lexington, KY 40506, USA},
Andrei Alexandru
  \addressmark[UK],
Ying Chen
  \address{Institute of High Energy Physics, 
           Academia Sinica, 
           Beijing 100039, P.R. China},
Shao-Jing Dong
  \addressmark[UK],
Ivan Horv\'{a}th
  \addressmark[UK],
Frank Lee
  \address{Center for Nuclear Studies, 
           Dept.\ of Physics, 
           George Washington Univ.,
           Washington, DC 20052, USA}
  \address{Jefferson Lab, 
           12000 Jefferson Avenue, 
           Newport News, VA 23606, USA},
Nilmani Mathur
  \addressmark[UK],
Harry B. Thacker
  \address{Department of Physics, 
           University of Virginia, 
           Charlottesville, VA 22901, USA},
Sonali Tamhankar
  \addressmark[UK],
Jianbo Zhang
\address{CSSM and Department of Physics,
         University of Adelaide, 
         Adelaide, SA 5005, Australia}
       }

\begin{abstract}
It is popular to probe the structure of the QCD vacuum indirectly by studying
individual fermion eigenmodes, because this provides a natural way to filter
out UV fluctuations.  The double-peaking in the distribution of the local
chiral orientation parameter ($X$) has been offered as evidence, by some, in
support of a particular model of the vacuum.  Here we caution that the
$X$-distribution peaking varies significantly with various versions of the
definition of $X$.  Furthermore, each distribution varies little from that
resulting from a random reshuffling of the left-handed (and independently the
right-handed) fields, which destroys any QCD-induced left-right correlation;
that is, the double-peaking is mostly a phase-space effect.  We propose a new
universal definition of the $X$ parameter whose distribution is uniform for
randomly reshuffled fields.  Any deviations from uniformity for actual data can
then be directly attributable to QCD-induced dynamics.  We find that the
familiar double peak disappears.
\end{abstract}

\maketitle


\section{Introduction}

The use of eigenmodes to probe the structure of the QCD vacuum is fairly old,
but the recent resurgence of interest is the result of the focus on local
properties~\cite{Horvath:2002} and the use of chiral
fermions~\cite{Neuberger:1998,Narayanan:1995}.  The space-time chiral structure
of the eigenmodes reflects the space-time topological structure of the
underlying gauge fields.  Our work~\cite{Us} has involved a detailed study of
the space-time geometric structure and we have argued that this is inconsistent
with the instanton liquid model (ILM).  Many others~\cite{Others} have studied
the behavior of the low lying eigenmodes and have argued in favor of the ILM,
but their conclusion relies solely on the interpretation of the observed
double-peaked behavior of the probability distribution of the local chiral
orientation parameter.  To help resolve this, here we revisit the local chiral
orientation parameter.


\section{Local Chirality Parameter}

The local chiral parameter, $X$, measures the tendency of a field at a
space-time point to be left or right handed.  It was
originally~\cite{Horvath:2002} defined in terms of the polar coordinate $\theta
= \tan^{-1} {|\psi_{x}^{L}|}/{|\psi_{x}^{R}|}$.  Many definitions of $X$ are
possible (including several from the literature~\cite{Others}):
\begin{eqnarray*} \label{alternate-X}
      X^{(1)}
& = & 
      \frac{4}{\pi}\tan^{-1} \frac{|\psi_{x}^{L}|}{|\psi_{x}^{R}|} - 1  \quad \equiv X \quad \\ 
      X^{(2)} 
& = & 
      \frac{4}{\pi}\tan^{-1} \frac{|\psi_{x}^{L}|^{2}}{|\psi_{x}^{R}|^{2}} - 1  \\ 
      X^{(1/2)} 
& = & 
      \frac{4}{\pi}\tan^{-1} \frac{|\psi_{x}^{L}|^{1/2}}{|\psi_{x}^{R}|^{1/2}} -1  \\ 
      X_{RBC} 
& = & 
      \frac{|\psi_{x}^{L}|^2-|\psi_{x}^{R}|^2}{|\psi_{x}^{L}|^2+|\psi_{x}^{R}|^2}
\end{eqnarray*} 
$X=+1 (-1)$ for a purely left (right) handed field.  However, as we see in
Fig.~{1}, the double peaking depends strongly on the definition of $X$.  We
need a universal measure which will produce a constant distribution for random,
uncorrelated data.

\begin{figure}[ht]
  \vspace{-0.9cm}
  \begin{center}
  \includegraphics[angle=0,width=\hsize]{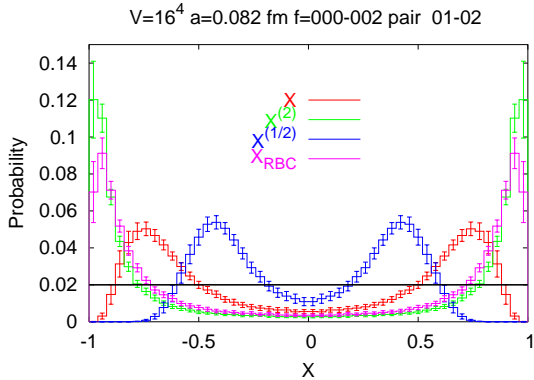}
  \vspace{-1.5cm}
  \caption{Alternate definitions of $X$.}
  \vspace{-1.5cm}
  \end{center}
\end{figure}
%


\section{A Test: Randomly Reshuffling Fields}

To determine how much of the double peaking in the $X$-distribution is due to
dynamics of the QCD vacuum, we make the following test: Compute the $X$
distribution from $\psi^{L}(x)$ and $\psi^{R}(x)$.  Then randomly reshuffle the
fields: $\psi^{L}_{\rm ran}(x) = \psi^L(x_{\rm ran})$, where $x_{\rm ran}$ is a
random permutation.  Use an independent random reshuffle for $\psi^{R}$.  Now
any QCD-dynamically-induced correlation between $\psi^{L}_{\rm ran}$ and
$\psi^{R}_{\rm ran}$ at site $x$ is destroyed.  Recompute the $X$-distribution.
We see in Fig.~{2} that the randomized $X$ distribution looks very similar.

\begin{figure}[h]
  \vspace{-0.9cm}
  \begin{center}
  \includegraphics[angle=0,width=\hsize]{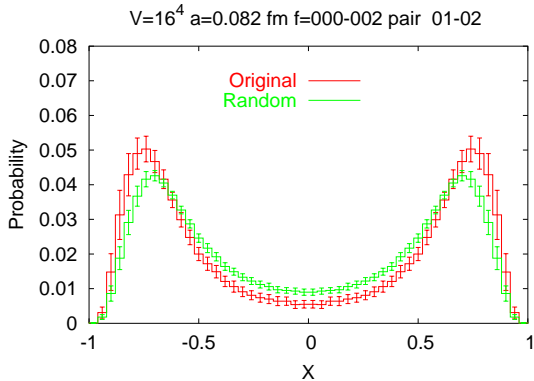}
  \vspace{-1.5cm}
  \caption{Original $X$-distribution and the $X$-distribution after random reshuffling.}
  \vspace{-0.5cm}
  \end{center}
\end{figure}

To understand how there can be so little difference between original and
randomized $X$-distributions, consider the marginal probability $p(\psi^{L}) =
\int d\psi^{R} \,\, p(\psi^{L},\psi^{R})$ of the actual (non-randomized) Monte
Carlo data (Fig.~{3}).  The key is that the marginal distribution has a long
tail.

\begin{figure}[h]
  \vspace{-0.9cm}
  \begin{center}
  \includegraphics[angle=0,width=\hsize]{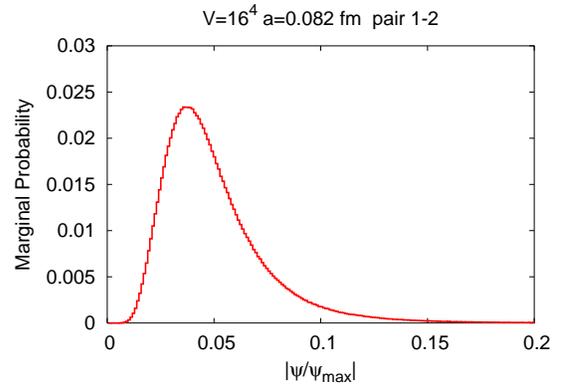}
  \vspace{-1.5cm}
  \caption{Marginal probability distribution for $|\psi^{L}|$ or $|\psi^{R}|$.}
  \vspace{-1.0cm}
  \end{center}
\end{figure}

To construct an $X$ distribution, one first cuts the data, selecting the most
intense points, i.e.\ the top 2\% (or 5\% or 10\%) of the density
$\psi^{\dagger}\psi= \psi^{L\dagger}\psi^{L}+ \psi^{R\dagger}\psi^{R}$.  For
$\psi^L$ and $\psi^R$ chosen randomly from this distribution, it is likely that
either $\psi^L$ or $\psi^R$ is large (to make the cut).  Given this
pre-selected data, and one of these large components (say $\psi^L$) in the tail
of the distribution, it is very likely that the randomly chosen $\psi^R$ is
much less than the selected $\psi^L$ (since the bulk of the marginal
distribution is in the peak at smaller values).  But such a pair (large
$\psi^L$ and smaller $\psi^R$, or vice versa) result in a value of $X$ much
different than $0$, leading to the ``double peak''.


\section{A New Universal Definition of $X$}

The ``$X$-distribution'' is a histogram of the probability $p(X)$ of a given
value of $X$.  With generic coordinates $q_1=|\psi^{R}|$ and $q_2=|\psi^{L}|$,
\[
      p(X) 
\, = \,
      \int dq_1 dq_2 \,\, p(q_1,q_2) \delta \left( X - X(q_1,q_2) \right)
\]
But if the components $q_1$ and $q_2$ were uncorrelated (supposing there were
no dynamics), then one would obtain a different histogram,
\[
      \tilde{p}(X) 
\, = \, 
      \int dq_1 dq_2 \,\, p(q_1) p(q_2) \delta \left( X - X(q_1,q_2) \right)
\]
Clearly, to expose the true correlation we need to ``subtract'' the phase space
background.  Consider the cumulative distribution function
$\tilde{P}(X)=\int_{-1}^{X} dX^{\prime} \,\, \tilde{p}(X^{\prime})$.  Then
$\tilde{P}\in[0,1]$.  For uncorrelated components, what is the probability
density of obtaining a particular value of $\tilde{P}$?
\begin{eqnarray*}
      \tilde{p}(\tilde{P}) 
      & = &
      \int_{-1}^{+1} dX \,\, \tilde{p}(X) 
      \delta\left(\tilde{P} - \int_{-1}^{X} dy \,\, \tilde{p}(y)\right) \\
      & = & 
      1
\end{eqnarray*}
That is, this probability density is uniform for uncorrelated data, which is
what we seek!  We can rescale and rename this quantity $\tilde{X}(X) = 2
\tilde{P}(X) - 1$ so that $\tilde{X} \in [-1,+1]$ as for $X$.  $\tilde{X}(X)$
is the improved quantity against which we should plot our distributions.  It
will be constant for uncorrelated data.  Deviations from uniformity will appear
for correlated data.


\section{Results}

\begin{figure}[ht]
  \vspace{-1.2cm}
  \begin{center}
  \includegraphics[angle=0,width=\hsize]{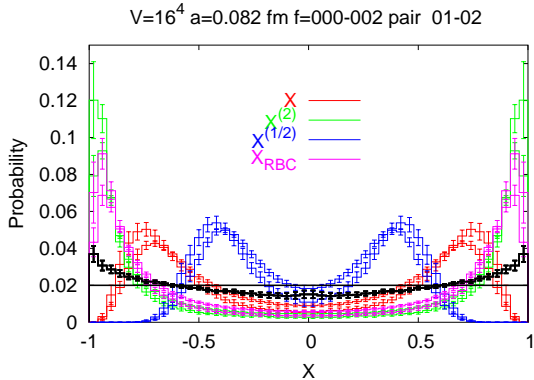}
  \vspace{-1.5cm}
  \caption{For various definitions of $X$: Original $X$-distribution.
           $X$-distribution after random reshuffle.  ``Difference'' (plot {\em
           vs.\/} new variable $\tilde{X}$).}
  \vspace{-1.0cm}
  \end{center}
\end{figure}

In Fig.~{4} for each of four definitions of $X$, we plot the original data and
its randomization versus $X$, and their ``difference'' (in thick black) versus
the new variable $\tilde{X}$.  Compare to the baseline of no correlation (thin
black line) at 0.02 (50 histogram bins).  Note that the new improved definition
$\tilde{X}$ is universal; that is, the black curves are exactly the same.  Thus
the ``Double-Peak'' is exposed as a phase-space effect, which is removed with
the new $\tilde{X}$.


\section{Summary}

Fermion eigenmodes (which filter UV fluctuations) are useful probes of the
structure of the QCD vacuum.  Observation of a double peaking in the local
chirality parameter (``$X$ distribution'') is a necessary but not sufficient
condition for the viability of the instanton liquid model (ILM)~\cite{Us}.  The
double-peaking depends strongly on the definition of $X$.  Moreover, after the
fields have been randomly reshuffled (thus destroying any QCD-dynamics
correlations), the $X$ distribution looks very similar to the original
distribution!  That is, most of the double peaking is due to phase space and
does not reflect the dynamics of QCD\@.  Accordingly, we have constructed a new
universal measure (``$\tilde{X}$'') of local chirality against which one should
plot probability distributions.  It is uniform for uncorrelated data.
Deviations from uniformity appear for correlated data.  The ``double-peak'',
exposed as mostly a phase-space effect, is flattened with the new improved
$\tilde{X}$, leaving little evidence in support of the ILM.


\end{document}